\documentclass[twocolumn,letter]{jpsj3} 

\usepackage{bm}
\usepackage{graphics}
\usepackage{graphicx}
\usepackage{eepic}
\usepackage{slashbox}
\title{Improvement on the GW$\Gamma$ Scheme for the Electron Self-Energy and\\ 
Relevance of the $G_0W_0$ Approximation from this Perspective} 

\author{Soh \textsc{Ishii}, Hideaki \textsc{Maebashi}$^{1}$ 
and Yasutami \textsc{Takada}$^{1}$}
\inst{Department of Physics, Graduate School of Engineering, 
Yokohama National University, \\ 
79-5 Tokiwadai, Hodogaya-ku, Yokohama, Kanagawa 240-8501 \\
$^{1}$Institute for Solid State Physics, University of Tokyo, 
Kashiwa, Chiba 277-8581
}

\abst{
Based on an exact functional form derived for the three-point vertex function 
$\Gamma$, we propose a self-consistent calculation scheme for the electron 
self-energy with $\Gamma$ always satisfying the Ward identity. This scheme is 
basically equivalent to the one proposed in 2001, but it is improved in the 
aspects of computational costs and its applicability range; it can treat a 
low-density electron system with a dielectric catastrophe. If it is applied to 
semiconductors and insulators, we find that the obtained quasiparticle dispersion 
is virtually the same as that in the one-shot $GW$ approximation (or $G_0W_0$A), 
indicating that the $G_0W_0$A actually takes proper account of both vertex and 
high-order self-energy corrections in a mutually cancelling manner. 
}
\kword{self-energy, $GW$ approximation, vertex correction, Ward identity, 
self-consistency, insulator, semiconductor, quasiparticle, Fermi liquid, 
Tomonaga-Luttinger liquid}

\def\runauthor{S. \textsc{Ishii}, H. \textsc{Maebashi} and Y. \textsc{Takada}
 }
\begin{document}
\maketitle
The electron self-energy $\Sigma$ is a fundamental quantity to control the 
quasiparticle properties in a many-electron system. Its accurate determination 
from first principles is recognized as a matter of central importance in many 
fields of condensed matter physics. In 1965, Hedin provided a nonperturbative 
self-consistent approach to $\Sigma$ in a closed set of equations, 
relating $\Sigma$ with the one-electron Green's function $G$, the dynamic screened 
interaction $W$, the polarization function $\Pi$, and the vertex function 
$\Gamma$~\cite{Hedin}. This formally exact formulation, however, allows of no 
practical implementation in its original form, because we cannot calculate the 
electron-hole irreducible interaction $\tilde{I}$, a key quantity in the 
Bethe-Salpeter equation to determine $\Gamma$, through its original 
definition using a functional derivative, $\tilde{I} \equiv \delta \Sigma/\delta G$. 
Thus we are compelled to adopt some approximate treatments such as the $GW$ 
approximation (GWA) in which $\Gamma$ is taken as unity. 

For more than two decades, successful calculations have been done for molecules, 
clusters, semiconductors, and insulators in the one-shot GWA (or 
$G_0W_0$A)~\cite{Hybertsen1,Hybertsen2,Aryasetiawan98,Aulbur00,Ishii,Kikuchi}, but 
this is usually regarded as a too primitive approximation, mostly because it is, 
in general, not a conserving approximation in the sense of Baym and 
Kadanoff~\cite{BK1,BK2}. In contrast, the GWA is a conserving one, obeying the 
conservation laws related to the macroscopic quantities like the total electron 
number. Upon implementation of this fully self-consistent GWA, however, we are 
led to a puzzling conclusion that the experiment on quasiparticle properties in 
semiconductors and insulators is much better described in the $G_0W_0$A than in 
the GWA~\cite{Eguiluz98,Delaney04}. A similar puzzle is also found in atoms and 
molecules~\cite{Stan09}. 

\begin{figure}[htbp]
\begin{center}
\includegraphics[scale=0.32,keepaspectratio]{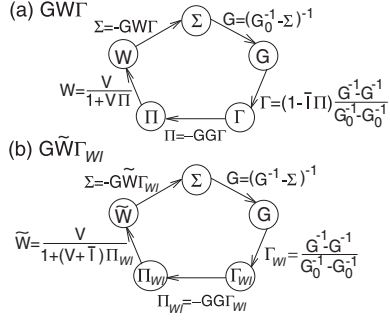}
\end{center}
\caption[Fig.1]{Iteration loops to determine the self-energy (a) in the original 
GW$\Gamma$ and (b) in the improved scheme, respectively.}
\label{fig:1}
\end{figure}

In metals, on the other hand, neither the $G_0W_0$A nor the GWA works very 
well~\cite{Kutepov09}, requiring us to include $\Gamma$ in some way in treating 
systems possessing gapless excitations. Some schemes have already been proposed 
for this purpose~\cite{Bruneval05,Shinshkin07}, but they do not satisfy the Ward 
identity (WI), an exact relation between $\Sigma$ and $\Gamma$ due to gauge 
invariance representing the local electron-number conservation~\cite{Takahashi57}. 
In 2001, based on general consideration on algorithms beyond the Baym-Kadanoff 
one~\cite{Takada95}, one of the authors (YT) proposed a scheme incorporating 
$\Gamma$ in the GWA with automatically fulfilling the WI~\cite{Takada01}. This 
GW$\Gamma$ scheme (see, Fig.~\ref{fig:1}(a)) succeeded in obtaining the correct 
quasiparticle behavior in simple metals, but it encounters a serious difficulty 
in the low-density electron gas; convergent results for $\Sigma$ are not 
obtained, if its density specified by the dimensionless parameter $r_s$ is 
larger than 5.25 where there appears the dielectric catastrophe associated with 
the divergence of the compressibility $\kappa$ at $r_s=5.25$ and concomitantly 
that of the static $\Pi$ in the long wave-length limit~\cite{Takada05,Maebashi09}. 
Incidentally the GWA does not suffer from this difficulty, because $\Pi$ in it 
is not a physical one satisfying the compressibility sum rule. 

In this Letter, we provide a new {\it exact} functional form for $\Gamma$, based 
on which we modify the GW$\Gamma$ into a scheme free from the difficulty 
originating from the dielectric catastrophe. In order to illustrate the power 
of the modified scheme, which will be referred to as G$\tilde{\rm W}\Gamma_{WI}$ 
(see, Fig.~\ref{fig:1}(b)), we show the results calculated for the electron gas 
at $r_s=8$. If it is applied to semiconductors and insulators, we find that 
the quasiparticle dispersion self-consistently obtained in the 
G$\tilde{\rm W}\Gamma_{WI}$ is essentially the same as that in the $G_0W_0$A, 
indicating that the $G_0W_0$A is superior to the GWA in the sense that for the 
systems with gapful excitations, it actually takes proper account of the mutual 
cancellation between vertex and high-order self-energy corrections. This 
observation resolves the above-mentioned long-standing puzzle on the GWA in 
comparison with the $G_0W_0$A. Here we emphasize that this cancellation is 
proved to be the case up to infinite order in an analytically rigorous way with 
clarifying the assumptions needed in the proof, in sharp contrast with the 
claims of a similar kind in the past;~\cite{DuBois1,DuBois2,Takada91} they were 
inferred from the behavior of low-order terms in perturbation expansion for metals. 

Let us start with recapitulating the exact relations for systems with 
translation symmetry in which momentum ${\bf p}$ is a good quantum number. 
The Dyson equation relates $G(p)$ with $\Sigma(p)$ through $G(p)^{-1} \!=\! 
{G_0(p)}^{-1}\!-\!\Sigma(p)$ with $p$ a combined notation of ${\bf p}$, spin 
$\sigma$, and fermion Matsubara frequency $i\omega_n \! \equiv \! i\pi T 
(2n\!+\!1)$ at temperature $T$ with an integer $n$~\cite{units}. The bare 
Green's function $G_0(p)$ is written as $G_0(p)\!=\!(i\omega_n\!-\!
\epsilon_{{\bf p}})^{-1}$ with $\epsilon_{{\bf p}}$ the bare one-electron 
dispersion. The Bethe-Salpeter equation determines $\Gamma(p,p\!+\!q)$ by 
\begin{eqnarray}
\label{eq:1}
\Gamma(p,p\!+\!q)&=& 1 + \sum_{p'} \tilde{I}(p,p\!+\!q;p',p'\!+\!q)
\nonumber \\
&&\times G(p')G(p'\!+\!q)\Gamma(p',p'\!+\!q), 
\end{eqnarray}
where $\sum_{p'}$ represents the sum $T\sum_{\omega_{n'}}\! \sum_{{\bf p'}
\sigma'}$. With use of $\Gamma(p,p\!+\!q)$, $\Pi(q)$ and $\Sigma(p)$ are, 
respectively, given by 
\begin{eqnarray}
\label{eq:2}
\Pi(q) &=& - \sum_{p}G(p)G(p\!+\!q)\Gamma(p,p\!+\!q),
\\
\label{eq:3}
\Sigma(p) &=&-\sum_{q}G(p\!+\!q)W(q)\Gamma(p,p\!+\!q),
\end{eqnarray}
with $W(q)=V({\bf q})/[1+V({\bf q})\Pi(q)]$, where $V({\bf q})$ is the bare 
Coulomb interaction $4\pi e^2/{\bf q}^2$. 

In Ref.~\citen{Takada01}, the concept of ``the ratio function'' was introduced 
to obtain an {\it approximate} functional form for $\Gamma(p,p\!+\!q)$ satisfying 
the WI. By exploiting this concept, we have explored an {\it exact} functional 
form for $\Gamma(p,p\!+\!q)$ and succeeded in obtaining the following form: 
\begin{eqnarray}
\label{eq:4a}
\Gamma(p,p\!+\!q)=\Gamma^{(a)}(p,p\!+\!q)\Gamma^{(b)}(p,p\!+\!q),
\end{eqnarray}
where $\Gamma^{(a)}(p,p\!+\!q)$ and $\Gamma^{(b)}(p,p\!+\!q)$ are, respectively, 
defined as 
$\Gamma^{(a)}(p,p\!+\!q)\!\equiv\!1\!-\!\langle \tilde{I} \rangle_{p,p+q}
\Pi(q)$ and 
\begin{eqnarray}
\label{eq:5}
\Gamma^{(b)}(p,p\!+\!q)  \!\equiv  \!
\frac{{G(p\!+\!q)}^{-1} \! - \! {G(p)}^{-1}} 
{{G_0(p\!+\!q)}^{-1}  \!- \! {G_0(p)}^{-1} \!- \!\Delta\Sigma_{p,p\!+\!q}}.
\end{eqnarray}
Here an average of $\tilde{I}$, $\langle \tilde{I} \rangle_{p,p+q}$, 
and a difference in the self-energy, $\Delta\Sigma_{p,p\!+\!q}$, are, 
respectively, introduced by
\begin{eqnarray}
\label{eq:4}
\langle \tilde{I} \rangle_{p,p+q} & \equiv &-\sum_{p'} 
\tilde{I}(p,p\!+\!q;p',p'\!+\!q)
\nonumber \\
&& \times G(p')G(p'\!+\!q)\Gamma(p',p'\!+\!q)/\Pi(q),
\end{eqnarray}
and 
\begin{eqnarray}
\label{eq:6}
\Delta\Sigma_{p,p\!+\!q} \! \equiv \! 
\sum_{p'}\!\tilde{I}(p,p\!+\!q;p',p'\!+\!q)[G(p'\!+\!q)\! - \! G(p')],
\end{eqnarray}
as functionals of $G$ and $\tilde{I}$. If $\tilde{I}$ is exact, 
$\Gamma^{(a)}(p,p\!+\!q)$ is nothing but $\Gamma(p,p\!+\!q)$ in Eq.~(\ref{eq:1}), 
as can easily be seen from the very definition of $\langle \tilde{I} 
\rangle_{p,p+q}$, and $\Delta\Sigma_{p,p\!+\!q}$ is reduced to 
$\Sigma(p\!+\!q)\!-\!\Sigma(p)$, leading to $\Gamma^{(b)}(p,p\!+\!q)\!=\!1$. 
Thus Eq.~(\ref{eq:4a}) provides the same $\Gamma(p,p\!+\!q)$ as that in the 
Hedin's exact theory. In reality, the exact $\tilde{I}$ is not known and we have 
to employ some approximate $\tilde{I}$, in which an advantage of Eq.~(\ref{eq:4a}) 
over Eq.~(\ref{eq:1}) becomes apparent; the former provides $\Gamma(p,p\!+\!q)$ 
satisfying the WI irrespective of the choice of $\tilde{I}$, while the latter 
does not. 

Physically $\tilde{I}$ takes care of exchange and correlation effects in 
$\Gamma(p,p\!+\!q)$ and it is well known that this physics can be captured 
by the local-field factor for the homogeneous electron gas or by the Jastrow 
factor for inhomogeneous systems. In either way, these effects are well described 
in terms of a function depending only on the inter-electron distance, which 
justifies to assume that $\tilde{I}(p,p\!+\!q;p',p'\!+\!q)$ depends only on $q$ 
to write $\tilde{I}(p,p\!+\!q;p',p'\!+\!q)=\bar{I}(q)$. If this assumption is 
adopted in our exact framework, we obtain $\langle \tilde{I} \rangle_{p,p+q}=
\bar{I}(q)$ and $\Delta\Sigma_{p,p\!+\!q}=0$. Then, by defining 
$\Gamma_{WI}(p,p\!+\!q)$ by 
\begin{eqnarray}
\label{eq:7}
\Gamma_{WI}(p,p\!+\!q) \equiv 
\frac{{G(p\!+\!q)}^{-1} \! - {G(p)}^{-1}}
{{G_0(p\!+\!q)}^{-1} - {G_0(p)}^{-1}}, 
\end{eqnarray}
we obtain $\Gamma(p,p\!+\!q)=[1-\bar{I}(q)\Pi(q)]\Gamma_{WI}(p,p\!+\!q)$, a 
result given in Ref.~\citen{Takada01}, leading to the GW$\Gamma$ in 
Fig.~\ref{fig:1}(a). 

By substituting this result of $\Gamma(p,p\!+\!q)$ into Eq.~(\ref{eq:2}), 
we find that $\Pi(q)$ is written as
\begin{eqnarray}
\label{eq:8}
\Pi(q) = \frac{\Pi_{WI}(q)}{1+\bar{I}(q) \Pi_{WI}(q)}
\end{eqnarray}
with $\Pi_{WI}(q)$, defined by
\begin{eqnarray}
\label{eq:9}
\Pi_{WI}(q) = - \sum_{p}G(p)G(p\!+\!q)\Gamma_{WI}(p,p\!+\!q).
\end{eqnarray}
Then we can rewrite $\Sigma(p)$ in Eq.~(\ref{eq:3}) into 
\begin{eqnarray}
\label{eq:10}
\Sigma(p) =-\sum_{q}G(p\!+\!q)\tilde{W}(q)\Gamma_{WI}(p,p\!+\!q),
\end{eqnarray}
with $\tilde{W}(q) \equiv V({\bf q})/\{1+[V({\bf q})+\bar{I}(q)]\Pi_{WI}(q)\}$. 
Combining these results, we can construct the G$\tilde{\rm W}\Gamma_{WI}$ scheme 
shown in Fig.~\ref{fig:1}(b). This scheme is equivalent to the GW$\Gamma$ 
in obtaining $\Sigma(p)$, but it is free from the problem of the dielectric 
catastrophe, because it does not contain the calculation of $\Pi(q)$ inside the 
iteration loop. 

It also renders a great advantage to the reduction of computational costs to 
calculate $\Pi(q)$ not directly but by Eq.~(\ref{eq:8}) via $\Pi_{WI}(q)$, because 
Eq.~(\ref{eq:9}) can be cast into a form convenient for numerical calculations as 
\begin{eqnarray}
\label{eq:11}
\Pi_{WI}(q) \equiv \Pi_{WI}({\bf q},i\omega_q) 
= \sum_{{\bf p}\sigma}\frac{n({\bf p\!+\!q})\!-\!n({\bf p})}
{i\omega_q \!-\!\epsilon_{{\bf p\!+\!q}}\!+\!\epsilon_{{\bf p}}},
\end{eqnarray}
where $\omega_q$ is the boson Matsubara frequency and  $n({\bf p})\,
[=T\sum_{\omega_{n}}\!G(p)e^{i\omega_n 0^+}]$ is 
the momentum distribution function. Note that this expression very much resembles 
the one for the polarization function in the random-phase approximation (RPA) 
$\Pi_{0}(q)$, which is given by
\begin{eqnarray}
\label{eq:12}
\Pi_{0}(q) \!=\! -\! \sum_{p}\!G_0(p)G_0(p\!+\!q)
\!=\! \sum_{{\bf p}\sigma}\!\frac{f(\epsilon_{{\bf p\!+\!q}})\!-\!
f(\epsilon_{{\bf p}})}
{i\omega_q \!-\!\epsilon_{{\bf p\!+\!q}}\!+\!\epsilon_{{\bf p}}},
\end{eqnarray}
where $f(\epsilon)$ is the Fermi distribution function. 

Two comments are in order: (i) Since it is conserved on the microscopic level 
in our scheme, the electron number is conserved on the macroscopic level as well. 
We can assure this conservation law by explicitly considering gauge invariance; 
because, as Baym discussed~\cite{BK2}, $G$ transforms in accord with $G_0$ 
with the change of gauge, $\Gamma_{WI}$ is gauge-invariant, implying that the 
conserving property of $\Sigma$ in the G$\tilde{\rm W}\Gamma_{WI}$ is the same 
as that without $\Gamma_{WI}$, {\it i.e.}, in the GWA. 
(ii) With use of Eq.~(\ref{eq:7}) and the introduction of ${\tilde \epsilon}_{\bf p} 
[\equiv\! \epsilon_{\bf p}\!-\!\sum_{q} \tilde{W}(q)/(i\omega_q \!-\!
\epsilon_{{\bf p\!+\!q}}\!+\!\epsilon_{{\bf p}})]$, our scheme provides an 
integral equation to determine $G(p)$ through 
\begin{eqnarray}
\label{eq:13}
(i \omega_n\!-\!{\tilde \epsilon}_{\bf p})G(p) = 1\!+\!\sum_{q}
\frac{\tilde{W}(q) G(p\!+\!q)}{i\omega_q \!-\!\epsilon_{{\bf p\!+\!q}}\!
+\!\epsilon_{{\bf p}}}\,.
\end{eqnarray}
On the assumption of $\bar{I}(q)=0$, this equation coincides with the one for 
obtaining the asymptotically exact $G(p)$ in a neutral Fermi system such as 
the one-dimensional Tomonaga-Luttinger model~\cite{DL74} or higher-dimensional 
models with strong forward scatterings~\cite{MCC98}. This coincidence clearly 
demonstrates the intrinsically nonperturbative nature of our framework.

Basically $\bar{I}(q)$ is at our disposal; it can be determined either 
by perturbation expansion or by some nonperturbative approach, but Eq.~(\ref{eq:8}) 
suggests us to choose $\bar{I}(q)=-G_{+}(q)V({\bf q})$ with $G_{+}(q)$ the 
local-field factor. Note, however, that the meaning of $G_{+}(q)$ here is different 
from the ordinary one that is defined with respect to $\Pi_{0}(q)$ 
instead of $\Pi_{WI}(q)$. Fortunately, we already know a good form for 
$G_{+}(q)$ with taking account of this subtle difference, which is $G_s(q)$ in 
Ref.~\citen{Richardson94}, satisfying the exact limit due to 
Niklasson~\cite{Niklasson74} as $|{\bf q}| \to \infty$. 

\begin{figure}[htbp]
\begin{center}
\includegraphics[scale=0.31,keepaspectratio]{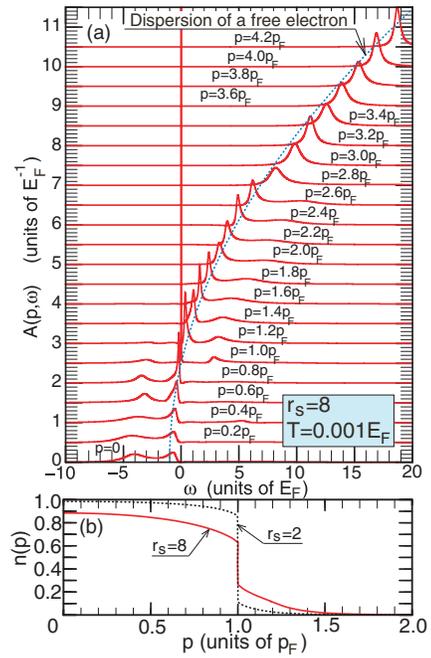}
\end{center}
\caption[Fig.2]{(a) One-electron spectral function $A({\bf p},\omega)$ and (b) 
momentum distribution function $n({\bf p})$ for the electron gas at $r_s=8$. 
For comparison, $n({\bf p})$ at $r_s=2$ is also shown.}
\label{fig:2}
\end{figure}

With this choice of $\bar{I}(q)$, the G$\tilde{\rm W}\Gamma_{WI}$ provides 
us the self-consistent $\Sigma(p)$ in the electron gas for $r_s>5.25$, in 
spite of the existence of the dielectric catastrophe associated with negative 
$\kappa$. After analytic continuation ($i\omega_n \to \omega \!+\! i0^+$) of 
$\Sigma(p)$ to the retarded self-energy $\Sigma^R({\bf p},\omega)$ with using 
the Pad\'{e} approximant, we obtain the one-electron spectral function 
$A({\bf p},\omega)\,[\equiv \!-{\rm Im}\, G^R({\bf p},\omega)/\pi]$; an example 
is plotted in Fig.~\ref{fig:2}(a) at $r_s\!=\!8$ and $T\!=\!0.001
E_{\rm F}$ with $E_{\rm F}$ the Fermi energy. The corresponding result for 
$n({\bf p})$ is given in Fig.~\ref{fig:2}(b), exhibiting a jump at the Fermi 
level, a typical Fermi-liquid property, though its deviation from $n_0({\bf p})
\,[\equiv \theta(p_{\rm F}\!-\!|{\bf p}|)]$ the step function with $p_{\rm F}$ 
the Fermi momentum is much larger than that at $r_s\!=\!2$, the typical density 
appropriate to many metals and semiconductors. We find an interesting result for 
the quasiparticle effective mass $m^*$ at $r_s=8$; for $|{\bf p}|\!<\!1.4p_{\rm F}$, 
$m^*$ is larger than $m_e$ the free-electron mass, implying dominance of the 
correlation effect over the exchange one, while the opposite is the case for 
$|{\bf p}|\!>\!1.4p_{\rm F}$ to give $m^*\!<\!m_e$. This crossover in $m^*$ never 
occurs for $r_s \leq 5$ where $m^*$ is always smaller than $m_e$~\cite{Takada01,Yasuhara99}. 

In the crystalline case, each quantity involved in the G$\tilde{\rm W}\Gamma_{WI}$ 
should be represented in the matrix form with respect to the reciprocal-lattice 
vectors $\{{\bf K}\}$. For example, $G(p)$ is a matrix composed of the elements 
$\{ G_{\bf K,K'}({\bf p},i\omega_n) \}$ with ${\bf p}$ a wave vector in the first 
Brillouin zone. For some quantities, we need to add the conversion factors 
transforming from the plane-wave basis to the Bloch-function one in considering 
the matrix elements; for example, ${\Pi_{0\,}}_{\bf K,K'}(q)$ is given as
\begin{eqnarray}
\label{eq:14}
&&\hskip -0.6truecm
{\Pi_{0\,}}_{\bf K,K'}(q) \!=\! \sum_{nn'{\bf p}\sigma}\!
\frac{f(\epsilon_{n'{\bf p\!+\!q}})\!-\!f(\epsilon_{n{\bf p}})}
{i\omega_q \!-\!\epsilon_{n'{\bf p\!+\!q}}\!+\!\epsilon_{n{\bf p}}}
\nonumber \\
&&\hskip -0.6truecm
\times \!
\langle n{\bf p}|e^{-i({\bf q\!+\!K})\cdot {\bf r}}|n'{\bf p\!+\!q}\rangle
\langle n'{\bf p\!+\!q}|e^{i({\bf q\!+\!K'})\cdot {\bf r'}}|n{\bf p}\rangle,
\end{eqnarray}
where $|n{\bf p}\rangle$ is the Bloch function for $n$th band~\cite{Adler62,Wiser63}. 

With this understanding, we have applied the G$\tilde{\rm W}\Gamma_{WI}$ to 
semiconductors and insulators possessing a gap in electronic excitation 
energies. Then, without detailed computations, the self-consistently 
determined quasiparticle energy $E_{{\bf p}}$ in our scheme is found to be 
well approximated by that in the $G_0W_0$A, as we explain in the following. 

Let us assume that $\Pi_{WI}(q)=\Pi_0(q)$ and $\bar{I}(q)=0$ for the time being. 
Then, we may rewrite Eq.~(\ref{eq:10}) as 
\begin{eqnarray}
\label{eq:15}
\Sigma(p) = 
-\!\sum_q \!\frac{W_0(q)}{G_0(p\!+\!q)^{-1} \!-\! G_0(p)^{-1}}
+ \lambda (p) G(p)^{-1},
\end{eqnarray}
with $W_0(q) \equiv V({\bf q})/[1+V({\bf q})\Pi_{0}(q)]$ and $\lambda (p)$ 
a dimensionless function, defined by
\begin{eqnarray}
\label{eq:16}
\lambda (p) \! \equiv \! \lambda ({\bf p}, i \omega_n) \!=\! \sum_q 
\frac{G(p+q)W_0(q)}{
i\omega_q \!-\!\epsilon_{{\bf p\!+\!q}}\!+\!\epsilon_{{\bf p}}}.
\end{eqnarray}
The quasiparticle dispersion $E_{{\bf p}}$ is determined by 
$G^R({\bf p},E_{{\bf p}})^{-1}=0$, amounting to $E_{{\bf p}}=\epsilon_{\bf p}+
\Sigma^R({\bf p},E_{{\bf p}})$, where we obtain the ``on-shell'' self-energy as 
\begin{eqnarray}
\label{eq:17}
\Sigma^R({\bf p},E_{{\bf p}}) = -\sum_q \frac{W_0(q)}
{i\omega_q \!-\!\epsilon_{{\bf p\!+\!q}}\!+\!\epsilon_{{\bf p}}}, 
\end{eqnarray}
by analytic continuation of $\Sigma(p)$ in Eq.~(\ref{eq:15}). In deriving 
Eq.~(\ref{eq:17}), we have paid due attention to the convergence of 
$\lambda^R ({\bf p},E_{{\bf p}})$ in gapful systems. In fact, provided that 
$\bar{I}(q)=0$, ${\tilde \epsilon}_{\bf p}$ and the integral in the right-hand 
side in Eq.~(\ref{eq:13}) are, respectively, reduced to $E_{{\bf p}}$ and 
$\lambda(p)$, leading to the behavior of $G^R({\bf p}, \omega)$ for $\omega$ 
near $E_{{\bf p}}$ as 
\begin{eqnarray}
\label{eq:18}
G^R({\bf p}, \omega) \approx 
\frac{1\!+\!\lambda^R ({\bf p}, E_{{\bf p}})}{\omega +i0^+- E_{{\bf p}}}. 
\end{eqnarray}

For comparison, let us consider the self-energy in the $G_0W_0$A, which is 
given by $\Sigma_0(p)\!=\!-\!\sum_q \!G_0(p\!+\!q)W_0(q)$. By analytic 
continuation $i \omega_n \to \epsilon_{\bf p}\!+\!i0^+$, we obtain 
\begin{eqnarray}
\label{eq:19}
&&
\hskip -1.0cm
\Sigma_0^R({\bf p},\epsilon_{\bf p}) \!=\!
-\!\sum_q \!\frac{W_0(q)}{i\omega_q \!-\!\epsilon_{{\bf p\!+\!q}}\!
+\!\epsilon_{\bf p}}\!-\! \frac{1}{2}\! \sum_{\bf q}\! 
W_0({\bf q}, \epsilon_{{\bf p\!+\!\bf q}} \!-\! \epsilon_{\bf p})
\nonumber\\ 
&&\times 
\Bigl [ \coth \frac{\epsilon_{{\bf p\!+\!\bf q}} \!-\! \epsilon_{\bf p}}{2T}
\!- \tanh \frac{\epsilon_{{\bf p\!+\!\bf q}}}{2T} \Bigr].
\end{eqnarray}
Because the transition ${\bf p\!+\!q}\to {\bf p}$ involved in Eq.~(\ref{eq:19}) 
is relevant only for the interband transition, $|\epsilon_{{\bf p\!+\!\bf q}} 
\!-\! \epsilon_{\bf p}|$ is always larger than $E_g$ the energy gap. At low $T$, 
the chemical potential $\mu$ lies at the center of the band gap, indicating that 
$|\epsilon_{{\bf p\!+\!\bf q}}| \!\ge\! E_g/2$. These two facts allow us to 
safely neglect the contribution from the second sum in Eq.~(\ref{eq:19}), as 
long as $T \ll E_g$. Thus we may write $E_{{\bf p}}^0$ the 
quasiparticle dispersion in the $G_0W_0$A as 
\begin{align}
\label{eq:20}
E_{{\bf p}}^0 = \epsilon_{{\bf p}} + 
\Sigma_0^R({\bf p},\epsilon_{{\bf p}}) 
=\epsilon_{{\bf p}} 
-\sum_q \frac{W_0(q)}
{i\omega_q \!-\!\epsilon_{{\bf p\!+\!q}}\!+\!\epsilon_{{\bf p}}},
\end{align}
leading us to conclude that $E_{{\bf p}}^0 = E_{{\bf p}}$. Note, however, that 
the spectral weight $z_{\bf p}\, [=(1\!-\!\partial \Sigma_0^R({\bf p},\omega)/
\partial \omega)^{-1}|_{\omega=\epsilon_{{\bf p}}}]$ is different from 
$1\!+\!\lambda^R ({\bf p}, E_{{\bf p}})$. 

In the literature, $E_{{\bf p}}^0$ is sometimes evaluated as $E_{{\bf p}}^0=
\epsilon_{{\bf p}}+z_{\bf p}\Sigma_0^R({\bf p},\epsilon_{{\bf p}})$ and there 
is a controversy as to whether this $z_{\bf p}$ should be included or not. 
As previously discussed in detail~\cite{Takada91}, we consider it better not 
to include $z_{\bf p}$ so that the vertex corrections beyond the RPA are 
properly included, together with higher-order self-energy terms in a mutually 
cancelling manner. In fact, our present result of $E_{{\bf p}}^0 = E_{{\bf p}}$ 
without this factor $z_{\bf p}$ indicates that this feature of mutual cancellation 
reaches far up to infinite order in semiconductors and insulators. 

Finally we comment on the two assumptions: (i) The difference between 
$\Pi_{WI}$ and $\Pi_0$ arises only from that between $n({\bf p})$ and 
$n_0({\bf p})$. In usual semiconductors and insulators, the valence-electron 
density is high; for example, $r_s=2$ for Si. Now $n({\bf p})$ in a metal at 
such $r_s$ does not deviate much from $n_0({\bf p})$ except for the 
states near the Fermi level, as shown in Fig.~\ref{fig:2}(b), but those states 
are absent from the outset in these gapful systems. Thus $n({\bf p})$ is 
close to $n_0({\bf p})$, leading to $\Pi_{WI} \approx \Pi_0$. 
(ii) Justification of $\bar{I}=0$ has already been done by numerical studies 
in Ref.~\citen{Hybertsen2}, in which $\bar{I}$ in our scheme is critically 
assessed in terms of $K_{xc}$ the density-derivative of the Kohn-Sham 
exchange-correlation potential. From an analytic 
point of view, it is enough to note that the basic processes to contribute to 
$\bar{I}$ are related to the interband electron-hole interactions, in which 
$|{\bf q}|$ for principal processes is of the order of $|{\bf K}|$, making 
$V({\bf q})$ very small and $G_{+}(q)$ reach its asymptotic constant. 
Thus the effect of $\bar{I}$ is weak in semiconductors and insulators. 

In summary, we have proposed the G$\tilde{\rm W}\Gamma_{WI}$ scheme for the 
fully self-consistent and conserving calculation of the electron self-energy. 
This can be applied not only to metals in a wide range of densities but also 
to semiconductors and insulators, in which the obtained 
quasiparticle dispersion is close to that in the $G_0W_0$A, explaining, 
from a fundamental viewpoint of many-body physics, the reason why the 
$G_0W_0$A better describes the experiment than the GWA in those gapful systems. 
We also realize that the role of $\bar{I}$, representing short-range exchange 
and correlation effects, is very much different between gapless and gapful 
systems; in the former, it can never be ignored to obtain the reliable quasiparticle 
behavior, but in the latter, it can be neglected, as long as the $G_0W_0$A well 
reproduces the experiment. In this respect we can suggest that the $G_0W_0$A 
should be performed with judiciously choosing the basis functions to make 
$\bar{I}$ and the difference between $\Pi_{WI}$ and $\Pi_0$ as small as possible. 

This work is supported 
by a Grant-in-Aid for Scientific Research (C) (No. 21540353) from MEXT, Japan. 


\end{document}